\documentclass[12pt]{article}
\usepackage{amssymb,latexsym,mathrsfs}
\usepackage{graphicx}
\usepackage{fancyhdr}
\usepackage[colorlinks=true, linkcolor=blue, citecolor=blue,urlcolor=blue, breaklinks=true]{hyperref}
 \usepackage{lineno}
\usepackage{sectsty}
\usepackage{array}
\usepackage{booktabs}

\begin{document}

\sectionfont{\large}
\subsectionfont{\normalsize}
\begin{center}
\textbf{\large Introduction to Delay Models and Their Wave Solutions} 
\end{center}

\normalsize
\begin{center}
Majid Bani-Yaghoub,\\
Department of Mathematics and Statistics\\
                University of Missouri-Kansas City\\
                Kansas City, Missouri  64110-2499, USA\\
								baniyaghoubm@umkc.edu\\
\end{center}

\normalsize
\vspace{15pt}
\begin{center}
\noindent\textbf{Abstract}
\end{center}
In this paper, a brief review of delay population models and their applications in ecology is provided. The inclusion of diffusion and nonlocality terms in delay models has given more capabilities to these models enabling them to capture several ecological phenomena such as the Allee effect, waves of invasive species and spatio-temporal competitions of interacting species. Moreover, recent advances in the studies of traveling and stationary wave solutions of delay models are outlined. In particular, the existence of stationary and traveling wave solutions of delay models, stability of wave solutions, formation of wavefronts in the special domain, and possible outcomes of delay models are discussed.    \\

\noindent \textbf{Math. Subj. Classification:} 37N25 (Dynamical systems in biology),  	37G15(Dynamical systems and ergodic theory )\\


\section{Mathematical Models with Delay}
\label{intro}
Ordinary and partial differential equations (ODEs and PDEs) have proven to be effective instruments in the study of real world problems in biology and medicine \cite{Bani08} 
\cite{Bani Miami} \cite{bani10}. They will unquestionably continue to serve as interdisciplinary tools in future research in mathematical biology. However, better comprehension of biological mechanisms demand for more sophisticated mathematical models. In particular, dynamics of infectious diseases and also population dynamics can be more realistically modeled  when time delays are integrated into the corresponding differential equations \cite{bani4} \cite{bani1} \cite{bani2} \cite{bani11} \cite{Kuang}.\\
\indent Indeed, the use of delay differential equations (DDEs) in modeling different biological situations has been on a rapid ascension \cite{baniC}. A simple example occurs due to the fact that the population density of a single species is directly dependent on the food resources. A shortage in the resources takes time to be resolved which can result in fluctuations in population growth or even in extinction of the entire population. Another example emerging in nature is reforestation. After replanting a hewn forest, it will take a minimum of twenty years for trees to reach any level of maturation. Hence, modeling any reforestation without considering delay terms is an approximation at best.\\
\indent Systems of ODEs and PDEs without delay are governed by the principle of causality; That is, "the future state of the system is independent of the past and is determined solely by the present" \cite{Kuang}. In various situations a model is realistic if it includes some of the past history of the system. Therefore, in systems of DDEs, the derivation at any given time depends on the solution at prior times.\\
\indent The history of DDEs applied in population biology goes back to predator-prey systems introduced by Italian mathematician Vito Volterra. In the 1920s he was asked to study the fluctuation observed in the fish population of the Adriatic Sea. In 1926 Volterra constructed the model \cite{Volterra 1926}.  Around the same time, the mathematical biologist A.J. Lotka, proposed a similar model but in a distinct context \cite{Lotka 1925}, which would later be recognized as the Lotka-Volterra model. These two investigators established their original work on the expression of predator-prey and competing relations in the form of nonlinear differential equations. The main goal of the Lotka-Volterra model is the quantification of the interaction of populations and also their inorganic environment and investigation of the temporal and spatial (if any) variation of members of various species.\\
\subsection{Spatially Homogeneous Delay Models}
\indent Perhaps the simplest type of past dependence is 
given by the discrete delay differential equation
\begin{equation}
\label{eq:basicRDDE}
\dot{x}(t)=f(t,x(t), x(t-\tau)),
\end{equation}
where the time delay $\tau$ is a positive constant.\\
A special case of (\ref{eq:basicRDDE}) is the well-known Hutchison's equation (1948) given by
\begin{equation}
\label{eq:HutchRDDE}
\dot{x}(t)=\gamma x(t)[1-x(t-\tau)/k],
\end{equation}
where $\gamma$, $k$ and $\tau >0$, \\
Equation (\ref{eq:HutchRDDE}) is also referred to as a delayed logistic equation (with a discrete delay). When the past dependence is through the derivative of the state variable, the equation is called a neutral functional differential equation (NFDE). An example of this is the following neutral delay logistic equation,
\begin{equation}
\label{eq:Gop-ZhangNFDE}
\dot{x}(t)=rx(t)\left[1-(x(t-\tau)+\rho\dot{x}(t-\tau))/k\right],
\end{equation}
which was first introduced by Gopalsamy and Zhang \cite{Gopal-Zhang}
in 1988. The global and local qualitative analysis of equation (\ref{eq:Gop-ZhangNFDE})  and also the boundedness of its solutions have been investigated in distinct studies \cite{FreedKuang 1991}, \cite{KuangFeld 1991}, \cite{FreedKuang 1992}. Based on equation (\ref{eq:Gop-ZhangNFDE}), the neutral predator-prey and neutral competition models were introduced and studied by Kuang \cite{Kuang 1992a}, \cite{Kuang 1992b}.\\
The following integro-differential equation is an extension to model (\ref{eq:HutchRDDE}) and is considered for the parasite population growth in the work of MacDonald (1978) \cite{MacDonald 1978}:
\begin{equation}
\label{eq:MacDpop}
\frac{dN(t)}{dt}=r N(t)\left[1-\frac{N(t)}{k}-\int^{t}_{0} N(s) G(t-s) ds\right],
\end{equation}
where $r$ and $k$ are positive, and instantaneous self-crowding term $G(t)$ is accompanied by a population term $N(t)$. It can be observed that the population $N(t)$ rises to a maximum and then exhibits an exponential decay. Equation (\ref {eq:MacDpop}) is originally taken from the work by Volterra's model for the effect of a deteriorating environment caused by the accumulation of waste products on mortality. In contrast with the commonly used equation (\ref{eq:HutchRDDE}) with discrete delay, equation (\ref {eq:MacDpop}) includes a type of delay that is known as distributed delay which is the sum of infinitely numerous small delays in the form of an integral. 
Although discrete delay can be appropriate in various situations \cite{6 blue} \cite{6 blue: 21}, \cite{6 blue: 22}, \cite{6 blue: 74}, the use of distributed delay allows us to include stochastic effects in a model \cite{6 blue: 11}, \cite{MacDonald 1978}. Otherwise the model is considered to be deterministic.\\
\indent One of the earliest delayed epidemic models was proposed by A.J. Lotka and F.R. Sharpe \cite{Sharpe Lotka 1923} in 1923. They examined the effect of incubation delays in a human mosquito model which was based on the malaria epidemics model by Ross \cite{Ross 1911}. Several delay models have been constructed to study the dynamics of infectious diseases such as Rabies, HIV, tuberculosis and influenza. In fact, a discrete delay can  be added to the model to account for the time lag between the moment a cell gets infected and the point when infected cells start producing virus. Such a delay is considered in some relatively recent works \cite{Murray I:371}, \cite{Murray I:372}, \cite{Murray I:407}, \cite{Murray I:408}, \cite{Murray I:440}, \cite{Murray I:441}, \cite{Murray I:226} thus furthering progress in correctly modeling the HIV infection process.\\
\indent In the absence of spatial variations, a Lotka-Volterra competition system \cite{Volterra 1931} with delay effects can be presented in the following couple of integro-differential equations
\begin{equation}
\label{eq1:LVmodel}
\dot{x}{(t)}=x{(t)}[a-bx(t)-\int^{0}_{-\tau}F_{1}(\theta)y(t+\theta)d\theta],
\end{equation}
\begin{equation}
\label{eq2:LVmodel}
\dot{y}{(t)}=y{(t)}[-\delta+cx(t)+\int^{0}_{-\tau}F_{2}(\theta)x(t+\theta)d\theta],
\end{equation}
where $x$ and $y$ are density of prey and predators, respectively and all constants and functions are nonnegative. The functions $F_{1}$  and  $F_{2}$ are delay kernels representing the manner in which the past history of predator and prey influence the current growth rates of prey and predator respectively.\\
In a more general manner, delayed predator-prey and competition models can take the form:
\begin{equation}
\label{eq1:genLVmodel}
\dot{x}{(t)}=x{(t)}F(t,x_{t},y_{t}),
\end{equation}
\begin{equation}
\label{eq2:genLVmodel}
\dot{y}{(t)}=y{(t)}G(t,x_{t},y_{t}),
\end{equation}
where the system state $x_{t}(\theta)=x(t+\theta)$, $y_{t}(\theta)=y(t+\theta)$ is defined for $\theta\leq0$ and $F$ and $G$ satisfy appropriate conditions such as  $\partial F/\partial y_{t}<0$, $\partial G/\partial x_{t}<0$, which represent a competition incorporated with delay. In other words, a past increase in the size of either populations $x$ or $y$ tends to a rapid decrease in the growth rate of the other population. \\
There have been several studies on system (\ref{eq1:LVmodel})-(\ref{eq2:LVmodel}) which is known as distributed delay Lotka-Volterra competition system. For instance, necessary and sufficient conditions for existence of positive periodic solutions of this system have been established in the work by Y. Li \cite{26 black}. Furthermore, global asymptotic stability and equilibrium coexistence are established in \cite{27 black}.\\

\subsection{Models with Delay and Diffusion}
\indent 
The celebrated Lotka-Volterra model for prey-predator dynamics has been generalized in diverse ways \cite{6 blue: 11}, \cite{May1973}, \cite{Murray1977}, \cite{Levin1978a}, \cite{Nicolis-prig 1977}. Spatial effects in either discrete or continuous, combination of diffusion and delays \cite{Xiao: 267}, internal population structure and delay with nonlocal effects \cite{Xiao: 97}, and interaction of several species are some of the extensions to prey-predator models. A rudimentary way to take into account the population spatial spread into delay models is to add a diffusion term into the system of DDEs. Then, without considering the interactions between delay and diffusion, a prey-predator model can often be presented by a reaction-diffusion (RD) system in the form,
\begin{equation}
\label{eq:ModelPreyPred1}
\frac{\partial u}{\partial t}(t,x)=d_{1}\Delta u(t,x)+ u(t,x)g(x,u(t,x)) -f_{1}(x,u(t,x),v(t,x)),\\
\end{equation}
\begin{equation}
\label{eq:ModelPreyPred}
\frac{\partial v}{\partial t}(t,x)=(d_{2}\Delta -\mu(x))v(t,x)-f_{2}(x,u(t-\tau,x),v(t-\tau,x)),
\end{equation}
where $v(t,x)$ and $u(t,x)$ denote the predator and prey biomass density respectively, $\mu(x)$ is the per-capita predator mortality rate at point $x$, $\tau$ is the constant delay (interpreted as the average time it takes to convert prey biomass into predator biomass), $d_{1}$ and $d_{2}$ are the corresponding diffusion coefficients and $f_{2}(x,u(t-\tau,x),v(t-\tau,x))$ is the biomass gain rate of predator at point $x$ and time $t$.\\
\indent The technical difficulty with this model arises when the diffusing predator that is at $x$ and time $t$ was not at $x$ at time $t-\tau$. Therefore, an interaction between delay and diffusion must be taken into account to incorporate the movements of the predator during the assimilation process. There are two main approaches to deal with this dilemma: the random-walk argument proposed by N.F. Briton \cite{6 blue: 8} and reduction of an age-structured model introduced by H. Smith and H. Thieme \cite{6 blue: 63}. 
Here, considering an age-structured modeling approach leads us to the following form of delayed RD equations with the nonlocality in the second equation that is caused by predators moving during biomass assimilation.
\begin{equation}
\label{eq:ModelPreyPred11}
\frac{\partial u}{\partial t}(t,x)=d_{1}\Delta u(t,x)+ u(t,x)g(x,u(t,x)) -f_{1}(x,u(t,x),v(t,x)),\\
\end{equation}
\begin{equation}
\label{eq:ageStruct RD}
\frac{\partial v}{\partial t}(t,x)=(d_{2}\Delta-\mu(x))v(t,x)
+\int_{\Omega} \Gamma_{2}(x,y,\tau)f_{2}(y,u(t-\tau,y), v(t-\tau,y)) dy,\\
\end{equation}
where $t >0$, $x\in \Omega$; $\Gamma_{2}$ is the appropriate Green's function of fundamental solution associated with $d\Delta_{2}-\mu(x)$ and possibly boundary conditions.\\
\indent Equation ({\ref{eq:ModelPreyPred11}) is a standard prey RD equation without time delay interacting with diffusion. In contrast, equation (\ref {eq:ageStruct RD}) is a predator RD equation with nonlinear reaction terms involving spatial nonlocality and time delay incorporated into the integral form. It can be observed that equation (\ref {eq:ageStruct RD}) is a differentiation of the integral equation,
\begin{equation}
\label{eq2:ageStruct RD} 
v(t,x)=\int^{t-\tau}_{-\infty}\left(\int_{\Omega}\Gamma(x,y,t-s)f(y,u(s,y),v(s,y)) dy\right) ds.
\end{equation}
Hence, in contrast to some other works \cite{Xiao: 97}, the nonlocal terms have been incorporated into the predator rather in the prey equation. The existence, uniqueness and boundedness of solutions of system (\ref{eq:ModelPreyPred11})-(\ref{eq:ageStruct RD}) were established in the work of Xiao-Qiang Zhao \cite{Zhao}. Moreover, despite the coexistence of prey and predator derived from the previous work \cite{Xiao: 97}, it is shown that the predator may become extinct if certain conditions are violated. In addition, the global attractivity of steady states of (\ref{eq:ageStruct RD}) is proved by the method fluctuation.\\
\indent An extended form of ({\ref{eq:ModelPreyPred11})-(\ref{eq:ageStruct RD}) has been recently proposed by C. Ou and J. Wu \cite{1 blue} that has attracted many researchers in the field of mathematical biology. The general delayed nonlocal RD system is presented by  
\begin{equation}
\label{eqCh3:nonlocalRD1}
\frac{\partial u(x,t)}{\partial t}=D\nabla^{2}u(x,t)+F(u(x,t),\int^{0}_{-\tau}\int_{\Omega} d\mu_{\tau}(\theta,y) g(u(x+y,t+\theta))),
\end{equation}
where $x\in\mathbb{R}^{m}$, $t\geq0$, $u(x,t)\in\mathbb{R}^{n}$, $D=diag(d_{1},...,d_{n})$ with positive constants $d_{i}$, $i=1,...,n$, $\tau>0$, $\Omega\subset\mathbb{R}^{m}$, $\mu_{\tau}:[-\tau,0]\times\Omega\rightarrow\mathbb{R}^{n\times n}$ is a normalized variation function (i.e. $\int^{0}_{-\tau}\int_{\Omega}d\mu_{\tau}(\theta,y)=1$), $F:\mathbb{R}^{n}\times\mathbb{R}^{n}\rightarrow\mathbb{R}^{n}$ and $g:\mathbb{R}^{n}\rightarrow\mathbb{R}^{n}$ are $C^{2}$-smooth functions.\\
The integral term in the function $F$ is a spatio-temporal convolution that represents in a very general form the interaction of time delay and spatial variations of individuals in a single species population. 

\section{Traveling Waveforms in Biology }

\indent A key element to a vast number of phenomena in biology is the appearance of a traveling wave in the spatial domain due to diffusion effects. Traveling waves of chemical concentrations, spread of pest outbreak, colonization of space by a population, spatial spread of epidemics, traveling waves in predator-prey systems, waves in excitable media, traveling frontal waves of a growing population and multispecies population dynamics with dispersal are all examples of the abundant studies that have been conducted in the area of biological waves. Several introductory and advanced books have included a large number of mathematical models that are related to different wave phenomena in biology (see for example \cite{Murray I},\cite{Murray II}, \cite{Murray I:584}, \cite{Murray I:487}, \cite{Murray I:45}, \cite{Murray I:201}, \cite{Okubo Bk}, \cite{Volpert bk}, \cite{Xiao: 267}). \\
\indent Systems of RD equations have been the main source in the study of various aspects of traveling wave behavior for decades. Although it has been recognized that some of the well-known results of traveling waves can be extended to delayed RD equations, in most of the cases, such extension becomes highly nontrivial. In particular, the equations describing waves are no longer systems of ODEs but rather DDEs. Yet, there has been substantial progress in the study of traveling waves in delayed RD systems (see for example \cite{1 blue}, \cite{1 blue: 47}, \cite{1 blue: 59}, \cite{7 blue}). In order to introduce some of the recent achievements \cite{bani12}, \cite{bani13}, first it is necessary to provide a pedagogical exposition of some basic theory applied in the study of traveling wave solutions of ordinary RD systems (i.e. RD systems with no delay). \\

\subsection{Wavefronts of Ordinary Reaction-Diffusion Systems}
\indent A traveling wave is usually taken to be a wave which travels without change of shape. To describe traveling waves mathematically, consider the following ordinary RD system
\begin{equation}
\label{eq1:waveRDsys}
\frac{du}{dt}=D\frac{\partial^{2}u}{\partial x^{2}}+f(u),
\end{equation}
where the vector $u(x,t)$ is for instance, the vector of different chemical concentrations at time $t$ and space $x\in\mathbb{R}^{m}$, $f(u)$ represents kinetics and $D$ is the diagonal matrix of diffusion coefficients. Then a solution $u(x,t)$ of (\ref{eq1:waveRDsys}) is a traveling wave if it is in the form of
\begin{equation}
\label{eq2:waveRDsys}
u(x,t)=U(\nu\cdot x-ct)=U(z),  z=x-ct, \mbox{ with }    x\in\mathbb{R} \mbox{ and }t>0,
\end{equation}
where $\nu$ is the unit vector in $\mathbb{R}^{m}$, $c$ is a constant that is called speed of propagation and the dependent variable $z$ is called the wave variable. Then $u(x,t)$ is a traveling wave moving at constant speed c in the $\nu$ direction without changing its shape. To be physically realistic, $u(z)$ must be bounded and nonnegative for all $z$. \\

\begin{figure}
\centering
  \includegraphics[width=0.75\textwidth]{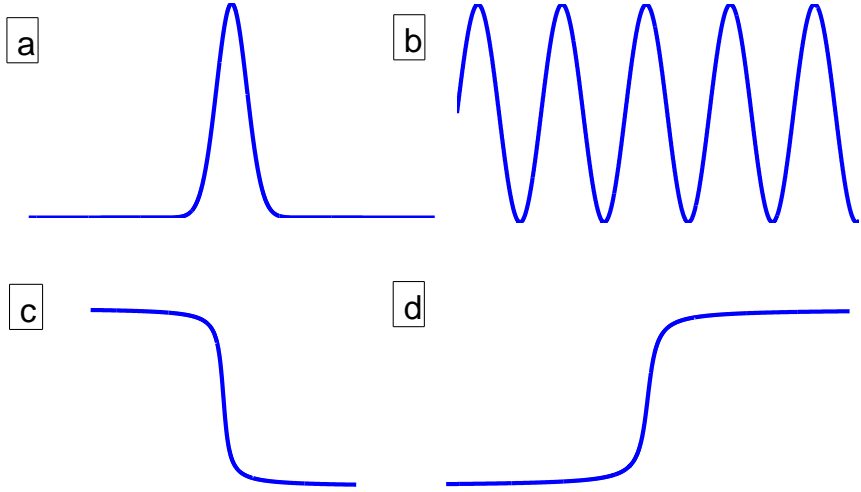}
\caption{\textit{Typical traveling wave solutions of RD systems: (a) pulse (b) traveling wave train (c) back (d) front: in the one dimensional case, these waves travel in the $x$-direction with speed $c$.}}
\label{fig:allwave}      
\end{figure}

\indent Figure \ref{fig:allwave} represents typical shapes of traveling waves that are observed in different studies. Traveling wavetrains (Figure 1b) are spatially-periodic traveling waves such that $U(z+L)=U(z)$ for all $z$ for some $L>0$. Front, back, pulse and inverted pulse are traveling waves that are asymptotically constant. In particular, they converge to homogeneous steady states $E_{1}$ and $E_{2}$ of  (\ref{eq1:waveRDsys}). For fronts (and backs) we have $E_{1}<E_{2}$ ($E_{1}>E_{2}$), $\lim_{z\rightarrow-\infty}U(z)=E_{1}$ and $\lim_{z\rightarrow+\infty}U(z)=E_{2}$, whereas pluses occur when $E=E_{1}=E_{2}$ and $\lim_{z\rightarrow\pm\infty}U(z)=E$.\\
Note that traveling wave fronts and backs perform almost the same quality and that general outcomes are satisfied for both of them. Similarly, an upside-down pulse (an inverted pulse) shares several properties with a regular pulse (Figure 1a). In addition to vast applications of traveling wavefronts in biology, various kinds of waves have been observed in chemical reactions (\cite{Bj: 99}, \cite{Volpert bk}, \cite{Bani II} \cite{Bani TH} and references herein) as well as nonlinear optics \cite{Bj: 1}, water waves \cite{Bj: 38}, \cite{Bj: 43}, \cite{Bj: 89}, gas dynamics \cite{Bj: 171} and solid mechanics \cite{Bj: 172}, \cite{Bj: 173}, \cite{Bj: 174}, \cite{Bj: 59}. Our primary concern in the present work is the traveling wavefronts. 
In order to obtain solutions of the form (\ref{eq2:waveRDsys}), let us substitute $U(z)$ into (\ref{eq1:waveRDsys}). Then we have:
\begin{equation}
\label{eq3:waveRDsys}
DU^{''}+cU^{'}+f(U)=0,
\end{equation}
where $'$ denotes the derivative with respect to $z$.\\
\indent Depending on the form of the reaction term $f(u)$ and also the initial condition $u(x,0)= u_{o}(x)$ on (\ref {eq3:waveRDsys}), a wave solution may or may not exist. For instance when (\ref{eq1:waveRDsys}) is scalar and $f(u)=0$, the solution of (\ref{eq3:waveRDsys}) is given by $u(z)=A+Be^{-\frac{cz}{d}}$ with $A$ and $B$ constant integration. Then by boundedness of $u(z)$, $B$ must be zero and hence $u(z)=A$ which is not a traveling wave, but is just a constant solution. Moreover, there is a minumum $c^{*}$ for the wave speed $c$. For values of $c$ less than the minimal speed $c^{*}$, the wave solution remains stationary (i.e. it does not move in any direction) or the traveling wave becomes physically unrealistic (i.e.  $u(z) < 0$ for some values of $z\in\mathbb{R}$). Moreover, the minimal speed $c^{*}$ is often considered to be greater than zero and the case $c=0$ describes standing waves that do not move at all. In general, the wave speed is dependent on the form of reaction terms in RD equations \cite{Murray I: 499}. As we will show later, in simple cases, wave speed  can be presented in terms of parameters associated with the reaction term $f(u)$ and the diffusion matrix $D$.\\

\indent Since analytical solutions of (\ref{eq1:waveRDsys}), in the case $f$ nonlinear, may not be available in most cases, the basic method of linearization is applied to study the traveling wave solutions. Suppose (\ref{eq1:waveRDsys}) has two homogeneous steady states $E_{1}$ and $E_{2}$; then a local solution of (\ref{eq3:waveRDsys}) around $E_{i}$ can be represented by $U=E_{i}+\widetilde{U}$. Substituting such a solution in (\ref{eq3:waveRDsys}) and dropping the nonlinear terms deduced from Taylor expansion gives the form
\begin{equation}
\label{eq4:waveRDsys}
D\widetilde{U}^{''}+c\widetilde{U}^{'}+J_{f(E_{i})}\widetilde{U}=0 \mbox{ for }i=1,2, 
\end{equation}
where $J_{f(E_{i})}$ is the Jacobian matrix of $f$ evaluated at $E_{i}$.\\
In order to have a nontrivial solution of the form $\widetilde{U}=e^{\lambda s}$, we must have
\begin{equation}
\label{eq5:waveRDsys}
\Lambda^{Ei}_{o}(\lambda):=det(D\lambda^{2}+c\lambda I+J_{f(E_{i})}=0,
\end{equation}
that is the characteristic equation at $E_{i}$, $i=1,2$. The following conditions are necessary for existence of a typical traveling wave front $U$ of ordinary RD system (\ref{eq1:waveRDsys}) which connects the two steady states $E_{1}$ and $E_{2}$ at two ends.
\begin{itemize}
	\item i) The steady states $E_{1}$ and $E_{2}$ are hyperbolic (i.e. $\Lambda^{E_{i}}_{o}(\lambda)=0$ has no pure imaginary root for $i=1,2$).
	\item ii) $\exists c^{*}_{o}>0$; $\forall c\geq c^{*}_{o}$ $\Lambda^{E_{1}}_{o}(\lambda)$ has a negative real root. Moreover, $E_{1}$ is a stable node (i.e. $\Lambda^{E_{1}}_{o}(\lambda)=0\Rightarrow Re(\lambda)<0)$.
	\item iii) For all $c\geq c^{*}$, equation (\ref{eq3:waveRDsys}) has a solution $U$ satisfying\\
	$\lim_{z\rightarrow-\infty} U(z)=E_{1}$, $\lim_{z\rightarrow\infty} U(z)=E_{2}$.
\end{itemize}
The minimal speed $c^{*}_{o}$ (\cite{bani1}) can be defined as follows:
\begin{equation}
\label{eq6:waveRDsys}
c^{*}_{o}=\inf\left\{c>0\mid\Lambda^{E_{1}}_{o}(\lambda)=0, \mbox{ has a negative real root}\right\}.
\end{equation}
To have a better understanding of traveling wave solutions of ordinary RD equations, consider the following two examples.\\
\par
\noindent \textbf{Example 1. A linear RD model for biological invasions.} Let $f(u)=\alpha u$ with $\alpha>0$ in equation (\ref{eq1:waveRDsys}); then we have a linear RD equation of the form,
\begin{equation}
\label{eq7:waveRDsys}
\frac{\partial u}{\partial t}=D\frac{\partial^{2}u}{\partial x^{2}}+\alpha u, \mbox{ with }\alpha>0 \mbox{ and } x\in\mathbb{R},
\end{equation}
which is a simple model of populations of exponential growth (Malthusian populations) with diffusion that is used for biological invasions such as muskrat dispersal in Europe \cite{Skellam 1951a}, invasion of nine-banded armadillo in the United States \cite{Humphrey 1974} and spread of larch casebearer in northern Idaho \cite{Long 1977}. The exact solution of (\ref{eq7:waveRDsys}) with initial condition $u(x,0)=M\delta(x)$ (i.e. for a population $u$ initially concentrated at origin $x=0$ and diffusing in an unbounded space) is given by
\begin{equation}
\label{eq8:waveRDsys}
u(x,t)=\frac{M}{2(\pi Dt)^{\frac{1}{2}}}\exp\left\{\alpha t-\frac{x^{2}}{4Dt}\right\}.
\end{equation}
In order to see how a wave solution travels with time, let $u= k= constant$ (i.e. consider a front of isoconcentration of the population). Then solve (\ref {eq8:waveRDsys}) with respect to $x/t$ \cite{Kendall 1948}. This gives the result
\begin{equation}
\label{eq9:waveRDsys}
x/t=\pm\left[4\alpha D-2Dt^{-1}\ln t-4Dt^{-1}\ln((2\pi D)^{\frac{1}{2}}k/M)\right]^{
\frac{1}{2}}.
\end{equation}
Then the asymptotic solution (i.e. as $t\rightarrow\infty$) of $u(x,t)=k$ is given by,
\begin{equation}
\label{eq10:waveRDsys}
x/t=\pm 2(\alpha D)^{\frac{1}{2}}=c.
\end{equation}
In other words the population wave front propagates outward from the initial disturbance with a speed ultimately equal to $2(\alpha D)^{\frac{1}{2}}$. \\
Further studies on the general solutions of (\ref {eq7:waveRDsys}) with the following Gaussian initial distribution
\begin{equation}
\label{eq11:waveRDsys}
u(x,0)=\frac{M}{(2\pi)^{\frac{1}{2}}\sigma}\exp(-x^{2}/2\sigma^{2}_{o}),
\end{equation}
where $\sigma^{2}_{o}$ is the variance, demonstrates that using an argument similar to that used above gives the same speed of propagation, $2(\alpha D)^{\frac{1}{2}}$ \cite{Cars-Jaegar 1959}.\\
\indent One might conclude that the same would be true for any initial condition, but it is only true if the amplitude of the initial condition possesses sufficient rapidity as $\left|x\right|$ tends to infinity \cite{Kendall 1948}; \cite{Mollison 1977b}; \cite{Needham 1992}. Although the linear relation (i.e. $f(u)=\alpha u)$ in (\ref{eq7:waveRDsys}) sounds realistic for many models in biology and ecology, the essential role of nonlinearity is crucial in a large variety of models. In fact, several researchers have developed modifications of the simple RD models that can afford better descriptions of more complicated problems (see for example \cite{Lewis-Kareiv 1993}, \cite{Shigesada 1995}, \cite{Sharov-Lieb 1998},\cite{Lewis 1996}, \cite{Murray 1986}). \\
\par
\noindent\textbf{Example 2. Fisher nonlinear RD equation.} Let us consider a nonlinear ordinary RD equation which was first proposed by Fisher \cite{Murray I:148} to study the spatial spread of a favored gene in a population
\begin{equation}
\label{eq1:FisherRD}
\frac{du}{dt}= \frac{\partial^{2}u}{\partial x^{2}} + u(1-u).
\end{equation}
Lamost immediately, this equation and its traveling wave solution was studied in the work by Kolmogoroff et al. in 1937 \cite{Murray I:287}. In order to demonstrate traveling wave solutions of the scalar one-dimensional Fisher-Kolmogoroff equation (\ref {eq1:FisherRD}), consider
\begin{equation}
\label{eq2:FisherRD}
u(x,t)=w(x-ct)=w(s), s=x-ct.
\end{equation}
Then we arrive to the wave profile equation
\begin{equation}
\label{eq3:FisherRD}
w^{''}+cw^{'}(s)+w(1-w)=0.
\end{equation}
Since (\ref{eq1:FisherRD}) is invariant if $x\rightarrow-x$, $c$ may be negative. To be specific, we assume $c\geq0$.\\
While equation (\ref {eq1:FisherRD}) has been a basis for a large number of different models such as spread of early farming in Europe \cite{Murray I: 5}, \cite{Murray I: 6}, gene culture and waves of advance \cite{Murray I: 14}, it has also received continuous attention in the study and search for analytical wave solutions \cite{Murray I:287}, \cite{Barakat}, \cite{Landahl}, \cite{Skellam 1973}, \cite{Hadeler-Rothe}, \cite{Kametaka}, \cite{Fife-Pele}, \cite{Kuno}. Yet, there is no general analytical solution available for either equation (\ref {eq1:FisherRD}) or (\ref {eq3:FisherRD}). Nevertheless, diverse methods can be applied to find an approximation of the solution of (\ref {eq1:FisherRD}) (see for example \cite{Rotenberg}, \cite{Rotenberg}). For the wave profile equation (\ref {eq3:FisherRD}), there is an exact solution for a particular $c$ greater than 2 (see \cite{Murray I} chapter 13). Moreover, the asymptotic solutions of (\ref {eq3:FisherRD}) for the case $0<\frac{1}{c^{2}}\ll 1$, can be obtained \cite{Murray 1984}, \cite{Kevorkian}, \cite{Canosa} by multiple scale and singular perturbation methods. Monotone iteration methods can also be used to construct an iterative wave solution of a generalized form of (\ref {eq3:FisherRD}) with discrete delay \cite{Xiao: 267}.\\
\indent The two steady states of (\ref{eq1:FisherRD}) are $E_{1}=0$ and $E_{2}=1$, where the wave solution $W(s)$ connects $E_{1}$ and $E_{2}$ as $s$ tends to $-\infty$ and $+\infty$ respectively.\\
Solving (\ref{eq5:waveRDsys}) for the eigenvalue $\lambda$ shows that $E_{2}$ is an unstable saddle. For $c>2$,  $E_{1}$ is a stable node where $c<2$ results in $E_{1}$ becoming a stable spiral.
Clearly the minimal speed $c^{*}_{o}$ defined by (\ref{eq6:waveRDsys}) is determined by $c^{*}_{o}=2$.\\
\indent By continuity arguments, there is a trajectory from $E_{2}$ to $E_{1}$ lying in the second quadrant (i.e. $w\geq0$ and $w^{'}\leq 0$) of the phase plane, with $0\leq w\leq 1$ for all wave speeds $c\geq c^{*}_{o}=2$. For $c<2$ there are trajectories connecting the unstable saddle $E_{2}$ to the stable spiral at origin (i.e. $E_{1})$ and hence they spiral around the origin as $z\rightarrow -\infty$. Therefore, in this case the traveling waves are unrealistic since $w(z)<0$ for some $z$.\\

\subsection{Wavefronts of Delayed Reaction-Diffusion Systems}
Several studies include traveling wave solutions of delayed RD equations without nonlocal effects. Namely, the existence, uniqueness and asymptotic behavior of the wave solutions corresponding to the delayed RD equation,
\begin{equation}
\label{eq7:Fisher-Kolm}
\frac{\partial u(x,t)}{\partial t}=\frac{\partial^{2}u(x,t)}{\partial x^{2}}+f(u(x,t),u(x,t-\tau)), 
\end{equation}
with $0\leq u(x,t)\leq1$ for $x\in\mathbb{R}$, $t > 0$, have been explicitly studied in the book by Wu \cite{Xiao: 267}, in which the method of phase plane (i.e. implementation of graph equation for trajectories), linearization, comparison of trajectories and their continuous dependence on wave speed $c$, the method of super-subsolution combined with some results already known from ODEs have been used to show existence and uniqueness of minimal speed $c^{*}_{\tau}$ and a nontrivial strictly increasing wave solution corresponding to (\ref{eq7:Fisher-Kolm}). Furthermore, an iteration scheme in conjunction with the fixed point theorem can be used to construct traveling waves in a system of RD equations of the form (\ref{eq7:Fisher-Kolm}). The methods mentioned above are some of the common methods employed to study traveling wave solutions of RD systems with discrete or distributed delay. 
Here, we bring the following example to illustrate the effect of delay on traveling waves. \\
\par
\noindent\textbf{Example 3. Delayed Fisher-Kolmogoroff RD equation.} Let $f = u(x,t-\tau)[1-u(x,t)]$ in the equation (\ref{eq7:Fisher-Kolm}); then we have
\begin{equation}
\label{eq1:Fisher-Kolm}
\frac{\partial u(x,t)}{\partial t}=\frac{\partial ^{2}u(x,t)}{\partial x^{2}}+u(x,t-\tau)[1-u(x,t)],
\end{equation}
which is also the natural extension of the logistic growth population model (\ref{eq:HutchRDDE}) and the population disperses via linear diffusion. Then the time delay $\tau>0$ represents the time required for the food supply (of the population) to recover from grazing (i.e. food supply at time $t$ depends on population size $u$ at time $t-\tau$).\\

The two steady states of (\ref{eq1:Fisher-Kolm}) are $E_{1}=k(\tau)=0$ and $E_{2}=1$ where $E_{2}$ is dependent on $\tau$. Using the following linear transformation we can obtain delay independent steady state $E_{1}$,

\begin{equation}
\widetilde{u}= \left\{
\begin{array}{ccc}
\label{eq2:Fisher-Kolm}
  \frac{u-E_{1}}{E_{2}-E_{1}} \mbox{ if } E_{1}\neq E_{2} \\  
  u-E_{1} \mbox{ otherwise .} 
  \end{array}
\right.
\end{equation}



For technical reasons, let $s=x+ct$ in (\ref {eq2:FisherRD}), then the wave profile equation corresponding to (\ref{eq1:Fisher-Kolm}),
\begin{equation}
\label{eq3:Fisher-Kolm}
w^{''}(s)-cw^{'}(s)+w(s-c\tau)[1-w(s)]=0,
\end{equation}
can be linearized around the new delay independent steady states $E_{1}$ and $E_{2}$ resulting in characteristic functions
\begin{equation}
\label{eq4:Fisher-Kolm}
\Lambda^{E_{i}}_{\tau}(\lambda):=\lambda^{2}-c\lambda+\alpha+\beta e^{-\lambda c\tau},
\end{equation}
where $\alpha$ and $\beta$ are correspondingly the partial derivatives of $w(s-c\tau)[1-w(s)]$ with respect to $w(s)$ and $w(s-c\tau)$ evaluated at $E_{i}$. Define
\begin{equation}
\label{eq5:Fisher-Kolm}
c^{*}_{\tau}:=\inf\left\{\ c>0;\mbox{ there is }s\in\mathbb{R}\mbox{ with }\Lambda^{E_{1}}_{\tau}(s)<0\right\}\;
\end{equation}
Then it can be shown that $c^{*}_{\tau}>0$ and for all $c>c^{*}_{\tau}$, there are exactly two real roots of $\Lambda^{E_{1}}_{\tau}$ such that $0<\lambda^{-}(c)<\lambda^{+}(c)<c$. Moreover, for $c=c^{*}_{\tau}$ there is exactly one real root $\lambda^{*}\in(0,c^{*})$. Since we let $s=x+ct$ and therefore the wave solution moves in the negative $x$-direction. The steady state $E_{1}$ is a stable node. In particular, it can be shown that for all $c> c^{*}$ there exists a unique (up-to-translation) nontrivial wave solution of (\ref {eq3:Fisher-Kolm}) with asymptotic behavior,
\begin{equation}
\label{eq6:Fisher-Kolm}
w_{c}(s)=ke^{\lambda^{-}(c)s}[1+0(1)]\mbox{ as }s\rightarrow-\infty,
\end{equation}
where $k$ is a constant.\\

\subsection{Periodic Traveling Wave Solutions}
\indent Periodic Traveling wave solutions (wavetrain solutions) of general RD systems with limit cycle kinetics have been widely studied. To introduce traveling wave trains, consider the system of RD equations
\begin{equation}
\label{eq1:wavetrRD}
\frac{\partial u}{\partial t}=D\nabla^{2}u+f(u).
\end{equation}
Assume that the spatially homogeneous system,
\begin{equation}
\label{eq2:wavetrRD}
\frac{dv}{dt}=f(v,\gamma),
\end{equation}
has a stable limit cycle when the bifurcation parameter $\gamma$ passes through the Hopf bifurcation value $\gamma_{c}$ (i.e. for $\gamma=\gamma_{c}+\epsilon$, $0<\epsilon\ll1$, a small amplitude limit cycle solution exists and is stable); then the RD system (\ref{eq1:wavetrRD}) admits a family of $2\pi$-periodic traveling  wavetrain solutions (originally called plane wave solutions) of the form
\begin{equation}
\label{eq3:wavetrRD}
u(x,t)=U(z), z=\sigma t-kx,
\end{equation}
with frequency $\sigma>0$, wavenumber $k$, wavelength $w=2\pi/k$ and speed $c=\sigma/k$.\\
\indent Existence and uniqueness of periodic traveling wave solutions (plane waves) of the form (\ref {eq3:wavetrRD}) is given in the seminal papers by Kopell and Howard \cite{Kopell2}, \cite{Kopell Howard}, \cite{Kopell Howard2}. A simpler way of discussing traveling wave solutions of this type and their applications can be found in the second volume by Murray \cite{Murray II}.\\

\indent The study of periodic traveling waves (sometimes called traveling oscillatory waves) of delayed RD systems has been the interest of much research. In particular, let $r\geq0$ given, $f: C([-r,0]; \mathbb{R}^{n})\rightarrow\mathbb{R}^{n}$ be a $C^{2}$ map, $D$ be an $N\times N$ real matrix with eigenvalues in the right-half complex plane. Consider the delayed RD system
\begin{equation}
\label{eq1:PeriodTrRD}
\frac{\partial}{\partial t}u(x,t)=D\Delta u(x,t)+f(u_{t}(x, \cdot)); x\in\mathbb{R}^{n}, t>0,
\end{equation}
where $x_{t}(x,\theta)=u(x,t+\theta)\mbox{ and }-r\leq\theta<0$.\\
\indent Then similar to what was mentioned above, system (\ref {eq1:PeriodTrRD}) admits a family of periodic traveling waves $U_{\theta}$ near the asymptotically stable $w$-periodic solution of $\dot{v}(t)=f(v_{t})$. Moreover, it can be shown that $U_{\theta}$ loses its stability due to perturbations by spatial diffusion (see \cite{Xiao: 267}; \cite{Oliveira}). Similarly destabilizing effects on spatially homogeneous periodic solutions of nonlinear diffusional models can be observed. For instance, the multi-scale perturbation method can be used to establish the existence of small amplitude traveling wavetrains of the following delayed diffusional predator-prey model \cite{Murray 1976}; Cohen et al. 1979}
\begin{equation}
\label{eq1:PerDiffModel}
\frac{\partial}{\partial t}u(x,t)=\frac{\partial^{2}}{\partial x^{2}}u(x,t)+\alpha\int^{\infty}_{o}k(s) u(x,s) ds+\epsilon(\int^{\infty}_{o}k(s) u(x,t-s) ds)^{3},
\end{equation}
with $0<\epsilon\ll1$.\\
\indent All small amplitude periodic wave solutions of (\ref {eq1:PerDiffModel}) are unstable solutions bifurcating from a spatially homogeneous steady state of the system (\ref {eq1:PerDiffModel}). There are several studies on the periodic traveling wave solutions of delayed diffusion models \cite{Bonilla}, \cite{Cohen-Rose}. Additionally, investigations on the qualitative behavior of  such solutions for nonlocal delayed RD systems have recently gained interest.

\section{Discussion}
In summary, we note that there have been widespread and extensive studies on delay models describing different biological situations. Incorporation of time delays into population models goes back to the works of Lotka and Volterra in the 1920s. However, the basic developments on the theory of DDEs were accomplished in the 1970s \cite{Green bk}, \cite{Hale 1993}, \cite{Kuang}. The local and global dynamics of DDEs arising from population biology, epidemiology and ecology are still of great interest \cite{Bani II} \cite{bani2}. In particular, there are intensive activities on wave solutions of delayed nonlocal diffusive systems \cite{bani12}  . Progress toward investigating the behaviors of such solutions may bring significant insights in understanding the nature of different biological phenomena. Moreover, mathematics has certainly become more productive through new techniques and methods constructed in analysis of traveling wave solutions as well as extensions to the existing methods. Thus far, the study of wave solutions of delayed RD systems has played a central role in understanding the qualitative behaviors of the solutions, but much more work needs to be done. Namely, despite substantial progress, the recent studies on wave fronts of delayed nonlocal RD equations (e.g. equation (\ref{eqCh3:nonlocalRD1}) are still in their early stages.

\end{document}